\documentstyle[12pt]{article}
\begin{document}
\title{ Information measures and classicality in quantum mechanics}
\author {C. Anastopoulos \\ Departament de Fisica Fonamental \\
Universitat de Barcelona \\ Av. Diagonal 647, 08028 Barcelona\\
Spain \\ 
 E-mail: charis@ffn.ub.es \\ UB/FFN/98-01}
\date { May 1997}
\maketitle
\begin{abstract}
We  study information measures in quantum mechanics, with the particular emphasis on providing a quantification of the notion of predictability and classicality. Our primary tool is the Shannon - Wehrl entropy $I$. We give a precise criterion for phase space classicality of states and argue that in view of this a) $I$ provides a good measure for the degree of deviation from classicality in closed systems, b) $I - S$ ($S$ the von Neumann entropy ) plays the same role in open quantum system. We examine particular examples in non-relativistic quantum mechanics. Finally we generalise  the discussion into the field theory case, and (this being one of our main motivations) we comment on the field classicalisation in early universe cosmology.
\end{abstract}
\pagebreak
 \renewcommand {\theequation}{\thesection.\arabic{equation}}
\let \ssection = \section
\renewcommand{\section}{\setcounter{equation}{0} \ssection}
\section{Introduction}
The subject of this paper is a discussion on the notion of classicality as emerging in quantum mechanical systems and the possibility of giving a quantification of the non-classical behaviour using an information theoretic measure. The physical context on which we mainly focus, is that of quantum fields in early universe cosmology.
\par
Recent years have seen a increased activity in the study of emergent classicality, which has  led to  the formation of new concepts and the significant increase in understanding the physical mechanisms underlying the classicalisation process. Key among the former are the notion of decoherence (either environmentally induced or through the intrinsic dynamics in closed systems) and its interplay with noise, setting limits to the degree of predictability enjoyed by any quantum mechanical system. As far as the latter is concerned, a large number of illustrative, exactly solvable models have been widely studied, mainly in the context of non - relativistic quantum mechanics.
\par
One of the driving forces of this activity, has been the need to understand the quantum to classical transition in a cosmological setting (quantum and early universe). In the context of the latter, it is well known that a basic premise of inflationary model is the eventual classicalisation of the quantum fluctuations as the seeds of later structure formation. Nevertheless in spite of the conceptual importance, 
it is fair to say that there is not yet a clear consensus on how the process of classicalisation is effected. The reason for this is partly that the well tested concepts have to be applied to a field theoretic setting with infinite number of degrees of freedom (hence besides the technical difficulties involved,  a postulated  split between system and environment is not intuitively transparent) and partly because of the fact that the relevant physics are somewhat remote from the better understood realm of the low energy world. By this, we mean that it is not easy to precisely identify, what is meant by classical behaviour and which physical quantities ought to exhibit it (is it the mean field \cite{CaHu2}, the field modes \cite{Star}, a coarse -grained version of the  n - pt functions \cite{CaHu2}?).
\par
We must also remark the absence from the discussion of a clear - cut and quantitatively precise criterion for classicality. Of course the formulation of such a criterion ought to depend upon the degrees of freedom one is  seeking to study. Very often there emergent characteristics of classicality are taken as definitive of it, something that might eventually lead to a confusion, as for instance when taking the large fluctuations as characteristic of classical behaviour.
The only general and unambiguous (provided one correctly identifies the relevant variables) criterion, is provided by the consistent histories approach to quantum mechanics \cite{Gri, Omn3, Omn, Har, GeHa}. Unfortunately the technical demands raised by this approach
are rather high, so that it has been possible to treat in detail only  a number of relatively simple systems. 
\par
The identification of a classicality criterion and the search of a measure to quantify it form the backbone of this paper. We argue that classicality ought to be though generically as a {\it phase space} manifestation \footnote{Note, that this does not preclude classicality emerging for much more coarse-grained variables, in particular at the level of hydrodynamics.}
and in that light the most suitable object for this task is a version of Shannon information: the Shannon -Wehrl (SW) entropy \cite{Weh, Wehr}. This has been considered before as a measure of quantum and environmentally induced fluctuations \cite{Hal, AnHa}.
\par
We expand on this previous work, by tying its properties with a precise formulation of a criterion of phase space classicality. The emergent criterion is influenced by the work of Omn\`es within the consistent histories program \cite{Omn}. It essentially states that a state is to be thought as classical if it is phase space concentrated and this property preserved by dynamical evolution. But here we apend an important distinction: classicality is destroyed not only in view of the increase of fluctuations but also because of phase space mixing induced by the quantum evolution. With few exceptions \cite{Omn2, Zur} this has not been focused properly in the existing bibliography and we proceed to examine it in detail. In particular we argue that large squeezing (typical for field modes in an expanding universe) is a characteristic of non - classical behaviour, something that is implicitly well known in the field of quantum optics.
\par
Our  criterion then is argued to entail that the SW entropy indeed quantifies the deviation from classicality : it takes into account both phase space spreading of the state as well as the phase space mixing. Hence it can be  translated in  the SW entropy taking values of the order of its lower bound of unity. We should stress the appealing fact that a single quantity is sufficient to capture the classicality relevant behaviour in even systems with a large number of degrees of freedom.
\par
The plan of the paper is then as follows. In the next section some preliminary definitions are given for information theory. We mainly focus on properties that are useful for the developement of our later argumentation. Section 3 is the main one. We give the definitions of the SW entropy, present some of its properties, state the classicality criterion and provide the connection between this and the SW entropy. A number of examples in non-relativistic quantum mechanics are studied so that particular features can be isolated and commented upon. In the next section the discussion is upgraded to the field theoretic context. Discussing the corresponding generalisations, we finally give a discussion of various proposals for field classicalisation as well as whether SW entropy  could  be identified with the phenomenological (thermodynamic) entropy, appearing in cosmological discussions.
\section{Shannon information in quantum mechanics}
\subsection{The notion of information}
Information is largely not an absolute concept. Intuitively it corresponds to the degree of precision of the knowledge we can have about a particular system.
As such, it has always to be defined with respect to the questions we want to ask. When one is dealing with systems exhibiting a degree of randomness, our knowledge about the is hidden in the assignment of probabilities to individual events.
\par
When one is dealing with alternatives that can be meaningfully assigned probabilities (either classical stochastic processes or quantum mechanics, but notably not quantum mechanical histories), one has an intuitive feeling of what properties a good measure of information should have: \\
1. It should be small for peaked probability distributions and large for spread ones ( reflecting the fact that there is less to be discovered by a measurement or a precise determination in the former case). \\
2. It should increase under coarse graining, i.e. when settling for a less detailed description of our system.
\\
These properties are nicely captured by Shannon's definition of information: Given a sample space $\Omega$ with $N$ elements and assignement of probabilities $p_i$ for $i \in \Omega$ then information is naturally defined as:
\begin{equation}
I_{\Omega}[p] = - \sum_i p_i \log p_i
\end{equation}
This is clearly a positive quantity, obtaining its maximum of $\log N$ for the total ignorance probability distribution  $p_i = 1/N$ and its minimum zero for a precise determination of an alternative. This incorporates nicely property 1), while property 2) is guaranteed by the concavity of this function. Hence, any coarse- graining $\Omega \rightarrow \Omega'$, with its corresponding restriction map for the probabilitites $p \rightarrow p'$ will entail
\begin{equation}
I_{\Omega}[p] \leq I_{\Omega'}[p']
\end{equation}
It is not our purpose to give an exhaustive list of the properties of the Shannon information here, since they are fully covered in the relevant bibliography\cite{Cov}. We just restrict ourselves to two important results.
\par
In the case of continuous sampling space $\Omega$ with a probability distribution $p(x) $ with $x \in \Omega$ the Shannon information is given by
\begin{equation}
I_{\Omega}[p] = - \int dx p(x) \log p(x)
\end{equation}
It is generically not positive and it may not even be bounded from below. In the case that $\Omega = R^n$ and for distributions with constant  covariance matrix $K$ it has a lower bound
\begin{equation}
I_{\Omega}[p] \geq 1 + \frac{1}{2} \log \det K
\end{equation}
which is achieved by the corresponding Gaussian probability distributions.
\par
Finally, we should note that one can define the relative information between two probability distributions $p_1$ and $p_2$ (henceforward we drop the subscript referring to sampling space unless explicitly required ) as
\begin{equation}
I[p_2|p_1] = \int dx p_1(x) (\log p_1(x) - \log p_2(x))
\end{equation}
This quantity is always positive and jointly convex with respect to both probability distributions. It is to be interpreted as the ``extra'' amount of information contained in $p_2$ with reference to $p_1$.

\subsection{The quantum mechanical context}
Quantum mechanics is  an inherently probabilistic theory. Given a quantum state $\rho$ one can construct probability measures for any observable by virtue of the spectral theorem.
\par
Shannon information can first be naturally defined with respect to any orthonormal basis on Hilbert space (hence with a maximal set of commuting observables). If we name the basis $|n \rangle$, then the probabilities $\langle n | \rho |n \rangle$ are constructed and the Shannon information $I_{\{n\}}[\rho]$ can be defined as in equation (2.1). Clearly the lower bound on $I$ is here again $0$
while the maximum bound is $\log N$ where $N$ is the dimension of the Hilbert space.
\par
Slightly more generally, one can define Shannon information with respect to any self- adjoint operator $A$ with discrete spectrum. Since then $A = \sum_n a_n P_n$, in term of the projectors $P_n$ in its eigenspaces we can define again 
\begin{equation}
I_A[\rho] = \sum_n tr (P_n \rho) \log tr(P_n \rho)
\end{equation}
In such a case the lower bound is not zero, unless $A$ has non-degenerate spectrum. This comes from the fact that in the degeneracy case, the probability distribution is a coarse graining with respect to the one defined by a maximal set of observables to which $A$ belongs.  
\par
There is an important relationship between Shannon information and von Neumann (vN) entropy $S[\rho] = - tr ( \rho \log \rho)$ in the case of discrete spectrum.
\begin{equation}
I_{\{n\}}[\rho] \geq S[\rho]
\end{equation}
The equality holds if  $\rho$ is diagonal in the $|n \rangle$ basis. Hence the values of the quantity $I _ S$ provide a measure of how close a density matrix is to be diagonal in the particular basis. This can be an important tool, in the context of environment superselection rules and the identification of the pointer basis.
\par
The case of continuous spectrum is rather more interesting. The projection valued measure $dE(x)$ associated to a self-adjoint operator $X$ defines a distribution function $p(x) = \frac{d}{dx} tr ( \rho E(x)$, with respect to which the Shannon information is defined. For the case of the position operator $x$ on $L^2(R)$ we get a lower bound for fixed uncertainty $\Delta x$ 
\begin{equation} 
I_x[\rho] \leq 1 + \log \left(2 \pi (\Delta x)^2   \right)
\end{equation}
saturated by the Gaussian states. 
A similar result holding for the momentum distribution $I_p[\rho]$ these can be combined with the standard uncertainty relation to yield
\begin{equation}
I_x[\rho] + I_p[\rho] \geq 1 + \log \pi \hbar
\end{equation}

\section{Shannon - Wehrl entropy}
When one needs to discuss the emergence of classical behaviour from a quantum system, one is in need to quantify the notion of fluctuations around classical predictability. In one dimensional case, the uncertainty $\Delta x \Delta p$ serves well this purpose, but in  systems with many degrees of freedom uncertainties are not by themselves sufficient to capture the classicalisation of the system's state. Correlations are involved (in the strong form of entanglement) that can disqualify even a localised in phase space state from being considered as classical. The same situation is of course more important in field theory, where one is working with infinite number of degrees of freedom.
\par
It is therefore important that simple quantities can be used to codify the classicality of a state. A particular variant of Shannon information, the so - called Shannon-Wehrl (SW) entropy, seems well suited to provide such a quantification. The purpose of this chapter is to explain in which sense the study of this object yields information about the classical behaviour of quantum states.
\par
Before proceeding we should be explicit in what we refer here as classicality. Two are the necessary requirements a state must satisfy in order to be charactrised as classical (or quasiclassical) \\
1. Suppression of interferences .\\
2. Seen as a wave packet, it has to evolve with a good degree of accuracy according to the classical equations f motion.
\\
The important point one has to stress here, is that we use the word classicality to refer to the Hamiltonian classical limit. Indeed (for instance in many body systems) classicality might refer to collective (hydrodynamic or thermodynamic) variables characterising the system. Our object is therefore concentrated on the phase space distributions associated with a quantum state. So suppression of interferences is implied with respect to some phase space ``basis'', an issue which is again very relevant when discussing classical equations of motion. Of course, given particular assumptions our discussion can involve classicality of collective variables as for instance center of mass of a many particle systems. For such issues, we refere the reader to \cite{Omn} for details. 
\par
 Therefore, classicality is defined with respect to {\it phase space properties}, rather than configuration space or momentum space ones. While phase space classicality straightforwardly implies configuration or momentum space one. The converse is not necessarily true.
 A state localised solely in position (and with a large momentum spread) cannot be considered as classical. The fluctuations around the classical path are too large, to destroy any sense of predictability. Moreover, such a localisation is not robust in the presence of even small interactions. 
\par
Finally, we  should remark that, since the SW entropy is defined in terms of coherent states, we have found expedient to employ intermittedly  the Schr\"oddinger and the Bargmann reperesentation, according to calculational convention.
\subsection{Definition and properties}
The SW entropy is defined as
\begin{equation}
I[\rho] = - \int Dw Dw^* p(w,w^*) \log p(w,w^*)
\end{equation}
in terms of the probability density
\begin{equation}
p(w,w^*) = \langle w|\rho| w \rangle
\end{equation}
where $|w \rangle$ is a (normalised) coherent state. Given the fact that $w$ is a complex linear combination of position and momentum (in the standard case of one dimensional harmonic oscillator  $ w = (\omega/2 \hbar)^{1/2} q + i (1/2 \hbar \omega)^{1/2} p$.) $p(w,w^*)$ can be viewed as a positive, normalised (due to the completeness relation of coherent states) distribution on phase space. This is invariably called the Q-symbol, or the Husimi distribution. It can be shown to correspond to a Gaussian smearing of the Wigner function (this rendering it positive).
\par
There is an ambiguity in the choice of the coherent states, essentially that they can be defined with respect to arbitrary state vector on the Hilbert space. Its resolution  by demanding that our information measure is shapest, will be dealt shortly. We just comment here, that standardly the coherent states can be taken as defined with respect to the vacuum of a harmonic oscillator or in the case of many dimensions of an isotropic harmonic oscillator. Then without loss of generality,  we can  represent $|w \rangle$ as
\begin{equation}
\langle x|w \rangle = \langle x| {\bf q p} \rangle = \left(2 \pi \hbar \sigma^2 \right)^{-1/4} \exp \left( -\frac{({\bf x}- {\bf q})^2}{4 \hbar \sigma^2} + i {\bf p} {\bf x}\right)
\end{equation}
 The SW entropy is the closest quantum object to the notion of Gibbs entropy (indeed Wehrl was calling it the classical entropy), in the sense that the coherent states define a cut-off phase space volume, with respect to which  a finite and unambiguous notion of entropy can be defined. Its lower bound is determined by two inequalities
\begin{eqnarray}
I[\rho] \geq S[\rho] \\
I[\rho] \geq 1
\end{eqnarray}
The latter is saturated by Gaussian coherent states (that it is only these states that achieve the minimum  is a non-trivial theorem due to Lieb \cite{Lie}), while the former by thermal states of harmonic oscillator in the high temperature regime.
\par
We should also remark that by  definition, the SW entropy of a state remains invariant when acting on the state with the elements of the Weyl group (translation in position and momentum).
\par
We should finally remark of an important property of the SW relative entropy. This is defined as
\begin{equation}
I[\rho_2|\rho_1] = \int Dw Dw^* p_{\rho_1}(w,w^*) \left( \log p_{\rho_1}(w,w^*) - \log p_{\rho_2}(w,w^*) \right)
\end{equation}
We have that
\begin{equation}
I[\rho_2|\rho_1] - I[\rho_2] + I [\rho_1] = \int Dw Dw^* (\log p_{\rho_1}(w,w^*) - \log p_{\rho_2}(w,w^*)) \leq 0
\end{equation}
since by construction $p_{\rho}(w,w^*) \leq 1$. Hence
\begin{equation}
I[\rho_2|\rho_1] \leq  I[\rho_2] - I [\rho_1]
\end{equation}
We are going to see later, that this inequality is saturated when $\rho_1$ is a coherent and $\rho_2$ a squeezed state with the same center. This property is not true for general probability distribution; in our case in holds by virtue of the particular definition of $p_{\rho}.$
\subsection{The classicality criterion}
The point we need to address now, is in which respect the SW entropy is a measure of phase space classicality, or put differently what is implied by the deviation of $I$ from its lower bound $I = 1$. This goes together with the resolution of the ambiguity, regarding the choice of coherent states in equation (3.2).
\par
The first point we need to stress, is that when one tries to give a phase space picture of quantum mechanical evolution and discsuss classicality, the need inevitably arises to introduce a measure of distance  on phase space. Indeed, in the simplest case of a single particle, to determine classicality (viewed as localisation) one is comparing the area in which the corresponding probability distribtion is supported with $\hbar$; indeed this criterion is encapsulated in the Heisenberg uncertainties: the phase space sampling has to be in a phase space shell of area much larger than $\hbar$, or correspondingly the state is viewed as classical if the uncertainty is of the order of magnitude of $\hbar$.
\par
In essence one needs to introduce a metric on the classical phase space. This is exactly, what a choice of a family of coherent states do. For, a coherent state is defined as $|pq \rangle = U(p,q) |\xi \rangle$, in terms of any vector of the Hilbert space. As such it defines a mapping $i_{\xi}$ from the phase space to the projective Hilbert space ${\cal RH}$. The latter is a K\"ahler manifold, thus having compatible metric and symplectic structures $g$ and $\Omega$. The pullbacks of these with respect to $i_{\xi}$ form the metric and symplectic structure respectively on the phase space. Hence, any choice of coherent state family defines a distinct metric on phase space with respect to which classicality is to be determined. The question then, translates into questioning which choice of metric is suitable for our purposes.
\par
The answer, is that this is largely irrelevant, provided some mild conditions are satisfied. First of all, we should note that there is an optimisation algorithm for coherent states of any group, so that the uncertainties  (or the determinant of the covariance matrix) of the relevant operators are minimal. In the standard case, this corresponds to defining coherent states with respect to the family of the Gaussians  ground states of some harmonic oscillator potential. But in fact, provided we take a sufficiently localised vector for $ |\xi \rangle$, this is not much of a restriction \footnote{ It is interesting to note, that at least in one approach to quantisation (Klauder's coherent state quantisation \cite{Kla}) a metric on the phase space is a primitive ingredient of the quantisation algorithm (so that the phase space can support Wiener measure). This could mean that that there is a preferred choice of an equivalence class of metrics, that give rise to unitarily equivalent  quantum theories. In the case of $L^2(R^n)$ these are the homogeneous metrics of zero curvature.}.
\par
The reason is mainly, that one of the important classicality criteria is the stability under time evolution. That is, a state is to be considered classical, if the determining criterion remains during its time evolution. This means that provided we have made a reasonable choice for our coherent state family, the object one should look is the relative information $I[\rho(0)|\rho(t)]$ where $\rho(t)$ is the evolved density matrix. This object is rather the one that should remain small, if the state $\rho$ is to be assessed as classical (provided of course that the peaks of the phase distribution approximately satisfy some deterministic equations of motion). Hence, the important criterion is eventually dynamical. We should choose a family of coherent states, that is rather stable with respect to time evolution. For harmonic oscillator potentials , this entails a particular choice for $| \xi \rangle$. But more generally, given the fact that the Gaussian approximation is good for a large class of potentials, it would be reasonable to consider the Gaussian coherent states for a larger class of systems. Alternatively, for highly non-linear potentials a good choice might be to take for $\xi$ lowest lying eigenstate of the Hamiltonian, even this being not a Gaussian. This becomes rather a necessity when one is dealaing with interacting field theory, as we shall xplain in the next chapter, and in general it seems wise when the Hamiltonian is invariant under  a group of symmetries, for they will be reflected in the choice of the metric. For similar reason, it seems more suitable to consider isotropic with respect to $q$ metrics for many dimensional systems.
\par
The question of course remains , what exactly is measured by the SW entropy. The answer, we will give here is simple: SW entropy is a measure of how much the 
``shape'' of the phase space distribution associated to a state $\rho$ deviates from the corresponding to coherent states. What we mean by shape, can be intuitively viewed in the Wigner function case. The $1 - \sigma$ contour of the Wigner function corresponding to a coherent state is a circle (the characterisation of circle follows from the choice of metric associated with this family of coherent states) with area $\hbar/2$. The SW entropy of a state $\rho$ is a quantification of the difference between this circle and the $1- \sigma$ contour associated with $\rho$. In particular, two characteristics are quantified: \\
1. The area enclosed in the contour. \\
2. The ``squeezing'' of the contour, i.e. the ratio of its length to its area (hence how much structure a state developes in the scale of $\hbar$). 
\par
In what follows, we shall try to explain both our interpretation of the SW entropy and its relevance for the classicality characterisation of a state. Later we shall give particular examples of our interpretation for the case of squeezed states.
\subsubsection{Phase space quasiprojectors}
One needs first to give a precise criterion for the notion of classicality of the state, and then examine how the use of the SW entropy, allows us to express this criterion in a quantified form.
\par
The approach we shall follow, is very much based on the ideas of Omn\'es \cite{Omn}, himself arguing within the context of the consistent histories approach to quantum mechanics. We believe his line of reasoning to allow for a sharp and precise characterisation of classicality. 
\par
In quantum mechanics one says that a state is localised with respect to some observable $A$ if it is an eigenstate of one of $A$ 's spectral projection. Actually approximate eigenstate is a sufficient characterisation. That is we can say that $\psi$ is localised in the range of the spectrum $[a,b]$ of $A$ if 
$|| E([a,b]) \psi - \psi || < \epsilon |b-a|$ for some  $\epsilon << 1$. (Note, a metric on the spectrum is implicitly assumed ).
\par
For the phase space localisation, one does not have projectors onto phase space ranges, but one can use rather unsharp phase space projectors (these are termed quasiprojectors by Omn\'es). These are essentially positive operator valued measures (POV) on the phase space \cite{Dav}, such that their marginal measures with respect to position and momentum space are respectively approximate position and momentum POV's.
\par
 To define such a family of quasiprojectors one needs first introduce a metric $g$ on phase space and its corresponding distance function $d$.
One can define the quasiprojector corresponding to a phase space cell 
$C$ through its Weyl symbol
\begin{equation}
f_C({\bf q}, {\bf p}) = \left(\frac{\hbar}{2 \pi}\right)^{n} \int d{\bf u} d{\bf v} e^{ i{\bf xu} + i {\bf p v}} Tr (e^{-i{\bf u} \hat{{\bf P}}- i {\bf v} \hat{{\bf Q}}} \hat{P}_C) 
\end{equation}
where  $2n$ the dimension of the phase space.  The Weyl symbol, ought to correspond to a smeared characteristic function. One can defined such one by considering for instant 
\begin{equation}
f_C({\bf q}, {\bf p} ) = \int_C \frac{d {\bf q'} d {\bf p'}}{(2 \pi \hbar)^n} 
\exp \left( - d^2[({\bf q},{\bf p}); ({\bf q'},{\bf p'})] \right)
\end{equation}
To each such projector one can associate a number $\epsilon$ which is roughly the ratio of volumes $[M]/[C]$. Here $[]$ stands for volume of a phase space cell and  $M$ is the margin of the phase space cell $C$, defined as the region where the smeared characteristic function of $C$ is appreciably different from 1 (well inside the cell) and 0 (outside the cell). If also, $\epsilon > e^{-l^2}$, where $l$ is the maximum curvature radius of the boundary if $C$, $P_C$ is close to a true projector, since the following properties are satisfied \\
1. $|P_C - P_C^2|_{tr} < c \epsilon |P_C|_{tr}$\\
2. if $C$ and $C'$ do not intersect $|P_C - P_{C'}|_{tr} < c' \epsilon \max (|P_C|_{tr},|P_{C'}|_{tr})$.
\\
with $c$ and $c'$ constants of order unity. Such phase space cells (regular in Omn\`es terminology), optimally have a value of $\epsilon$ of the order of $(\hbar/[C])^{n/2}$. For a given family of quasiprojectors ( meaning in particular a choice of metric on the phase space), we can view the optimal choice of $\epsilon$ for each cell as a function from the measurable phase space cells to the real positive numbers. We shall call it the {\it classicality function} associated with this choice of metric.
\subsubsection{The classicality criterion}
 Given a family of quasiprojectors, one can say that a state $|\psi \rangle$ is localised within a phase space cell $C$, if it is an $\epsilon$-approximate eigenstate of $P_C$, i.e. if 
\begin{equation}
||P_c |\psi \rangle - |\psi \rangle || \leq \epsilon
\end{equation}
But, we should remark, that localisation in phase space does not imply classicality. As we have seen localisation in phase space is relative to a choice of a phase space metric. Hence it is not a stringent criterion of classicality, let alone that one needs still ensure that the state remains localised during its classical evolution. This is an essential requirement, that largely removes the redundancy due to the freedom of choosing a phase space metric.
 \par
Hence a classicality criterion ought to read as follows: \\
\\
A pure state $\psi$ is considered to exhibit classical behaviour in some time interval $I$, if with respect to some choice of a family of quasiprojectors, is $\epsilon$- localised in phase space cells $C_t$, such that \\
1. $C_t$ is correlated  $C_{t`}$ by the classical equations of motion \\
2. $[C_t] << [C_I]$, where $C_I$ is the smallest $\epsilon$ - regular phase space cell that contains  the union of all $C_t$ 's.
\\ \\
The second condition is added here, so that time evolution is not trivial, i.e. there is indeed some meaning to a coarse grained description of the classical equations of motion. Another important point stressed here, is that classicality is contingent upon a particular time interval, outside of which phenomena of wave packet spreading might invalidate the localisation condition. This is even apparent for the case of free particle evolution, as we shall examine later.
\par
The above criterion was for pure states. In the case of mixed states (and relaxing the condition for unitary evolution), it should be generalised to include density matrices. This is straightforwardly done by substituting the approximate eigenstate condition (3.11) by the requirement 
\begin{equation}
|| P_C \rho P_C - \rho ||_{tr} \leq \epsilon
\end{equation}
the rest following as before.
\par
The above definitions contain nothing more than the intuitive idea, that a state is classical if its Wigner function exhibits a number of sufficiently concentrated peaks, each of which follows with some degree of approximation the classical equations of motion. Such a criterion has been widely used in the literature. The point we insist is rather the importance of the introduction of the metric in the phase space, as the one determining localisation. While intuitive arguments, based on uncertainty principle might usually be sufficient for the  determination of classicality. But what one may overlook in such considerations (particularly when one is dealing with many dimensional or field systems) is the loss of predictability (i.e the large growth of fluctuations) due  extreme squeezing in some directions of the phase space distribution. Such a phenomenon will generically cause the state not to be an approximate eigenfunction of the relevant  phase space projector, thus invalidating our criterion of classicality. This is particularly true in recent discussions on classicalisation of cosmological quantum fields \cite{Star}. 
\par
It is exactly at this point that the SW entropy proves to be a meaningful calculational tool. For as we measured it does not only measure the spread of the Wigner function, but also its shape. In other words, SW entropy measures the degree of approximation of the classical equations of motion to the Ehrenfest's theorem  mean values. The time parameter associated with the increase of the SW entropy (whenever this increases) , is essentially the parameter determining the breakdown of the classical approximation.
\par
Finally we should remark that a choice of coherent states defines naturally a family of quasiprojectors by
\begin{equation}
P_C = \int \frac{d {\bf p} d {\bf q}}{(2 \pi \hbar)^n} | {\bf qp} \rangle \langle {\bf qp}|
\end{equation}
These have actually been used by Omn\'es in the context of consistent histories to prove a semiclassical theorem. In view of our previous discussion, we can remark that computing the relative information between an initial coherent state and the evolved state at time $t$ provides a good measure of how much a particular Hamiltonian preserves or degrades classical predictability.

\subsubsection{Estimating the SW entropy} 
Before examining some concrete examples, we should at first examine how the phase space spread of  a state $\psi$ is encoded in the SW entropy. 
\par
Let us consider first the case of $\psi$ being an approximate eigenstate of a phase space projector $P_C$ with classicality parameter $\epsilon$. The probabilty distribution asscociated to $P_C$ , namely $\langle z|P_c|z \rangle / Tr P_C$ is within an approximation of $\epsilon$ a characteristic function of $C$ divided by the trace. But this is also a smearing of the distribution function corresponding to its eigenstate $\psi$. Hence due to the concavity of the entropy we have:
\begin{equation}
I[\psi] \leq \log ( Tr P_C) + O(\epsilon) = \log \frac{[C]}{(2 \pi \hbar)^n} + O(\epsilon)
\end{equation}
which is essentially the number of ``classical states'' within phase space volume $C$. Reasoning inversely if for a state $\psi$ (due for instance to time evolution) its SW entropy becomes much larger than $\log \frac{[C]}{(2 \pi \hbar)^n}$, its corresponding classicality parameter for its time evolution grows essentially as fast as $I[\psi]$ hence becoming of the order or larger than unity. 

\subsubsection{Linear canonical transformations}
 
The evaluation of SW and relative SW entropies for states that are obtained from implementing a linear canonical transformation on coherent state.
is quite important for a number of reasons. First, it gives an intuitive example of the way
entropy is connected with defomation of the shape of the $1 - \sigma$ contour. Second, this type of transformation appears naturally in time evolution of physically interesting systems: Hamiltonian evolution in the Gaussian approximation and in particular quantum fields in non-static spacetimes. Our results in this section will be valid for a description of such cases.
\par
Recall, that given a family of (Gaussian) coherent states $|w \rangle$ on some Hilbert space $L^2(R^n)$ the annihilation operators are naturally defined by
\begin{equation}
\hat{a}(\xi) |w \rangle = \xi^*_i w^i |w \rangle
\end{equation}
with $\xi, w \in C^n$ and can be written as $\hat{a}(\xi)r
 = \hat{a}^i \xi^*_i$.
\par
A linear canonical transformation is implemented by a unitary operator $S = e^{iA}$ where $A$ is a self-adjoint  quadratic to $\hat{a}$ and $\hat{a}^{\dagger}$ 
\begin{equation}
S \hat{a}(\xi) S^{-1} = \hat{a}(A^{\dagger} \xi) + \hat{a}^{\dagger}(B^{\dagger} \xi)
\end{equation}
where $A$ and $B$ are $n \times n$ complex matrices to be viewed as linear operators on the underlying real vector space $R^{2n}$. They are the parameters of the squeezing transformation and preservation of the canonical commutation reltions  enforces the Bogolubov identities:
\begin{eqnarray}
A^{\dagger}A - B^{\dagger}B = 1 \\
A A^{\dagger} - B B^{\dagger} = 1 \\
A^{\dagger} \bar{B}  = B^{\dagger} \bar{A}
\end{eqnarray}
where we use the bar to denote complex conjugation of a matrix. It is well known that the set of these transformations forms a reperesentation on our Hilbert space of the symplectic group $Sp(2n,R)$. Transformations with $B = 0$ are sometimes denoted as rotations ( forming a $U(n)$ subgroup) and ones generated by operators $A$ not containing terms mixing $a$ and $a^{\dagger}$ as squeezing.
 It is straightforward to check that the matrix elements of $S$ in a coherent state basis are given by
\begin{eqnarray}
\langle z|S|w \rangle = \left( \det(1-\bar{K}K) \right)^{-1/4} \nonumber \\ 
\exp \left(- |z|^2/2 - |w|^2/2 + \frac{1}{2} z^* K z^* + \frac{1}{2} w \bar{K} w  + z^* A^{-1} w \right)
\end{eqnarray}
Here $K$ stands for the matrix
\begin{equation}
K = A^{ -1} \bar{B}
\end{equation}
\par
The transformed  vacuum $|0;A,B \rangle$ is defined by the action of $S$ on $|0 \rangle$ and a transformed state $|w;A,B \rangle$ by the action of the operator $U(w)$ of the Weyl group on $|0;A,B \rangle$. Since the SW entropy is invariant under phase space translation, one can use the transformed  vacuum for its calculation.
\par
The corresponding probability distribution is
\begin{eqnarray}
p_{0;A,B}(w,w^*) = \left( \det(1 - \bar{K}K) \right)^{-1/2} \nonumber \\
\exp \left(  - |w|^2
+ \frac{1}{2} w^* K w^* + \frac{1}{2} w \bar{K}w \right)
\end{eqnarray}
From this one gets the following expression for the SW entropy of a transformed state (for brevity just use $A$ and $B$ as arguments)
\begin{equation}
I[A,B ] = 1 + \log [\det(1 - \bar{K}K)]^{-1/2} = 1 - \frac{1}{2} Tr \log (1 - \bar{K}K)
\end{equation}
Let us examine some special illustrative cases.
\\ \\
{\bf Pure rotation:} In such a case $K = 0$ and hence $I[A,0] = 1$ (the transformed state is a coherent state).
\\
\\ 
{\bf One dimensional case:} The general squeezing transformation is of the form
\begin{equation}
S \hat{a} S^{-1} = \cosh r \hat{a} - e^{i \phi} \sinh r \hat{a}^{\dagger}
\end{equation} 
in terms of the positive real $r$ and the phase $\phi$.
In this case $K = - e^{i \phi} \tanh r$ and the SW entropy for the squeezed states reads
\begin{equation}
I[r,\phi] = 1 + \log \cosh r
\end{equation}
This illustrates our earlier arguments since $r$ is interpreted as the eccentricity of the ellipse corresponding to the $1- \sigma$ contour of the squeezed state Wigner function. For large $r$ the ellipse becomes extremely prolongated in a direction determined by $\phi$ and the SW entropy grows linearly with $r$.
\\
\\
{\bf Two - mode squeezing:} There is a 6 - parameter family of squeezing transformation in two dimensions. A widely studied case is the  Caves - Schumaker squeezing, well studied in the field of quantum optics. This is generated by the unitary operator
\begin{equation}
S = \exp \left( r e^{i \phi} \hat{a}^{\dagger}_1 \hat{a}^{\dagger}_2 - r e^{-i \phi} \hat{a}_1 \hat{a}_2 \right) 
\end{equation}
and corresponds to the matrices
\begin{eqnarray}
A=  \left(\begin{array}{cc}
                 \cosh r & 0 \\        
                  0     & \cosh r \end{array} \right)
B = \left( \begin{array}{cc}
                 0 & - \sinh r e^{i \phi} \\        
                 - \sinh r e^{i \phi}& 0 \end{array} \right)
\end{eqnarray}
which yield the value 
\begin{equation}
I[r, \phi] = 1 + 2 \log \cosh r
\end{equation}
Note that here the parameter $r$ has a different physical interpretation. If our system represents two (non-identical) one dimensional particles, then the parameter $r$ is a measure of the entanglement of the total state. This is a non-classical feature; if our classical limit is to correspond to two classical particles the entanglement between them must be minimal. Hence SW entropy can quantify also this deviation from classicality (provided of course that the coherent state family with respect to which it is defined, is constructed from a factorised vacuum state).

\subsubsection{Relative entropy}
It is interesting also to compute the relative SW entropy between a coherent and a transformed state. Without loss of generality one can consider the coherent state to be the vacuum.
\par
The probability distribution associated to the transformed state $|z;A,B \rangle $ is
\begin{eqnarray}
p_{z;A,B}(w,w^*) = \left( \det(1 - \bar{K}K) \right)^{-1/2} \nonumber \\
\exp \left(  - |w-z|^2
+ \frac{1}{2} (w^* - z^*) K (w^* - z^*) + \frac{1}{2} (w - z) \bar{K}(w - z) \right)
\end{eqnarray}
hence the relative entropy with respect to vacuum is
\begin{equation}
I[0|z;A,B] =  \log [\det(1 - \bar{K}K)]^{-1/2} + |z|^2 + \frac{1}{2} (z^* K z^* + z \bar{K}z)
\end{equation}
 Hence the relative entropy is a sum of a term, purely from the squeezing plus a term containing the contribution of the Weyl translation. Note that in the case of $z = 0$ (pure linear tansformation) the inequality (3.8) is saturated. It is a reasonable conjecture, that this is true only for this particular class of states, i.e. for Gaussians with the same center. 
\subsection{Squeezing induced by quantum evolution}

\subsubsection{The Gaussian approximation}

We now come back to our main point. We shall consider the evolution of SW entropy for closed quantum systems , their evolution governed by a Hamiltonian $H = {\bf p}^2/2m + V({\bf q})$, in the Gaussian approximation. The latter consists essentially in approximating the evolution of Gaussian states, by the action of a linear canonical transformation. This is of course exact for systems evolving under a quadratic Hamiltonian and a good approximation for systems evolving in a macroscopically varying potential (at least within a particular time interval while the spread of the wave function has not become extremely large. We shall see that the evaluation of the SW entropy gives a self - consistency check for the validity o the Gaussian approximation.
\par
In this section, it is more convenient to switch back to the Schr\"oddinger representation for our Hilbert space vectors. Choosing our coherent state basis by
the relation
\begin{equation}
w^i = (2 \hbar \sigma^2)^{-1/2} q^i + i (\sigma^2/2 \hbar)^{1/2} p^i
\end{equation}
i.e. choosing an isotropic and factorised Gaussian defining state, we get the following expression for a translated vector $\psi$:
\begin{eqnarray}
\psi({\bf x}) = \left( \det M^* M {2 \pi \hbar} \right)^{-1/4} \nonumber \\
\exp \left( - \frac{1}{4 \hbar} (x-q)^i (LM^{-1})_{ij} (x-q)^j + i\ \hbar p_i x^i \right)
\end{eqnarray}
where the matrices $R$ and $S$ are such that $ \hbar^{1/2} MM^*$ and $ \hbar^{1/2}/2 L^*L$ are  the position and momentum covariance matrices respectively. Since the expression is invariant under a $U(n)$ matrix right acting on both $L$ and $M$ we have the freedom to define them in terms of the matrix $K$ of equation (3.21)  through the following relations
\begin{eqnarray}
Q = LM^{-1} = (2 \sigma^2)^{-1} (1+K)(1-K)^{-1} \\
M = (Re Q)^{-1/2} \\
L = QM
\end{eqnarray}
or what is more important the inverse relationship
\begin{equation}
K = (1 + Q)^{-1} (1 - Q)
\end{equation}
\par
Now, for the Gaussian approximation we shall utilise a result of Hagedorn \cite{Hag}. For a large class of physically relevant potentials (bounded from below, growing slower than a Gaussian) and  time interval $[0,T]$, the Gaussian (3.32) evolves to another Gaussian of the same type with the center determined by the classical equation of motion, a phase given by the corresponding classical action  and the matrices $A(t)$, $B(t)$ evolving according to the equations
\begin{eqnarray} 
\frac{d}{dt}A(t) = \frac{i}{2m} B(t) \\
\frac{d}{dt}B(t) = 2 i V^{(2)}(q(t)) A(t)
\end{eqnarray}
In fact they can be shown to satisfy
\begin{eqnarray}
M(t) = \frac{\partial q(t)}{\partial q(0)} M(0) + \frac{i}{2} \frac{\partial q(t)}{\partial p(0)} L(0)
\\
L(t) =  \frac{\partial p(t)}{\partial p(0)} L(0) - 2 i \frac{\partial p(t)}{\partial q(0)} M(0)
\end{eqnarray}
\par
To study the SW entropy production, we will consider the evolution of an initial coherent state (hence $K(0) = 0$) so that $M(0) = (2 \sigma^2)^{1/2} 1$ and $L(0) = (2 \sigma^2)^{-1/2} 1$.

\subsubsection{One dimensional case}
In the case of a free  particle the complex numbers $M$ and $L$ read 
\begin{eqnarray}
M(t) = (2 \sigma^2)^{1/2} + \frac{i}{2m} (2 \sigma^2)^{-1/2} t \\
L(t) = (2 \sigma^2) ^{-1/2}
\end{eqnarray}
Using the equations (3.39), (3.40) we find that  the SW entropy at large times ($ t >> \sigma^2 m $  behaves like
\begin{equation}
I \simeq 1 + \log \frac{t}{8 \sigma^2 m}
\end{equation}
 In a free particle time evolution produces strong squeezing, towards localising a particle in the position momentum (actually for free evolution momentum basis is some sort of pointer basis since superposition of two states with different momenta are asymptotically suppressed - though not exponentially as in the presence of environment). Hence eventually classical predictability breaks down for the free particle, though rather slowly. In view of our previous discussion the classicality parameter $\epsilon$ is increasing logarithmically with time. More precisely taking into account   equation (3.14) a state that can be considered as localised in a volume $V$ of phase space initially will stop being localised (approximate eigenstate of the corresponding quasiprojector) after time 
\begin{equation}
t \simeq \frac{4 \sigma^2 mV}{ \pi \hbar}
\end{equation}
Note the persistence of predictability for higher mass particles.
\par
For a harmonic oscillator Hamiltonian the choice of the standard coherent states with $ \sigma^2 = (4m \omega)^{-1}$ gives constant SW entropy, corresponding to maximum possible predictability. It is nonetheless instructive to see what would happen, had we not been wise to make this choice. The relevant quantities read now
\begin{eqnarray}
M(t) = (2 \sigma^2 \cos \omega t + \frac{i}{2m \omega} (2 \sigma^2)^{1/2} \sin \omega t \\
L(t) = (2 \sigma^2)^{-1/2} \cos \omega t - 2 i m \omega (2 \sigma^2)^{1/2} \sin \omega t
\end{eqnarray} 
It is easy now to verify that the SW entropy remains bounded for all times, taking values around 
$ 1 + |\log (4 m \omega \sigma^2)|$. Hence, the harmonic oscillator potential generally preserves classicality for large class of phase space localised states, the SW entropy remaining bounded by the limit  for the proper choice of the quasiprojectors monitoring the classical evolution).
 This also verifies that our classicality estimation based on the SW entropy is sufficiently stable with respect to the choice of coherent state family. 
\par
Of interest is also the case of the inverse harmonic oscillator potential. In the context of inflationary cosmology, it is sometimes states that a number of modes that evolve for a time as inverse harmonic oscillators, undergo amplification of their fluctuations and hence become ``classicalised''. As mentioned in the introduction, we believe that in such claims there is a confusion between the notion of large fluctuations and classicality. Amplified quantum mechanical fluctuations are {\it not} classical fluctuations. They only imply lack of predictability, which can be a purely quantum mechanical phenomenon \footnote{To see this it is sufficient to compute the time evolution of a superposition of two spatially localised states under this potentials. There is no way one could interpret their amplified fluctuations as classical however large they might become.}. For a potential $V(q) = - \frac{1}{2} m k^2 q^2$, it is easy to verify that for coherent states defined by $ \sigma^2 = (4m k)^{-1}$ the evolution is as squeezing in equation (3.24) with parameters $r = kt $ and $\phi = \pi/2$. Hence the SW entropy evolves as 
\begin{equation}
 I(t) = 1 + \log \cosh kt
\end{equation}
and asymptotically grows with  $kt$. 
\par
In the case of general  potentials in one dimension, one can make some qualitative predictions. If $V(q)$ is also bounded from above by $V_m$, for particles with $E >> V_m$ the results of the free particle case ought to be relevant: degradation of predictability growing logarithmically with time. For $U$- shaped potentials and low energies (hence mimicking a harmonic oscillator) predictability ought to remain good. Rugged potentials that vary within microscopic scales rapidly destroy predictability (tunnelling effects plus caustics typically weaken the effectiveness of the Gaussian approximation). 

\subsubsection{Higher dimensional systems}
\par
When considering systems with many degrees of freedom, equations (3.39), (3.40) can be the basis of a number of qualitative estimations. For instance, if the classical equations of motion  admit runaway solutions (positive Lyapunov exponents), the matrices $M$ and $L$ are going to have exponentially increasing with time entries and typically a behaviour of the type of inverse harmonic oscillator is to be expected.
\par
Hence, the SW entropy is going to increase linearly at long times, eventually bringing again a breakdown of classicality. Hence we arrive at a conclusion, noted by many authors, that quantum mechanical systems the classical analogue of which exhibits chaotic behaviour (meaning exponential dependence on the initial conditions), typically does not have a good classical limit \cite{Omn}. It is interesting to note that the SW entropy plays a role of a measure of mixing. By this we mean, the thin spreading of an initial phase space distribution into a given partition of phase space, so that its components eventually occupy larger and larger number of partition cells.  This suggests that the SW entropy plays a role similar to the classical Kolmogorov - Sinai entropy of classical dynamical systems (of course not sharing its invariance properties) and related measures of mixing.
\par
In view of our inequality (3.8)  the difference between SW entropies at initial time and time $t$ can be viewed as an estimation of the upper bound to the relative entropy between the classically evolved state ( Weyl transforming according to classical equation of motion) and full quantum evolution.
\par
A further remark is at point here. Classicality and in particular predictability is a ``non-perturbative'' issue. Even in the Gaussian approximation knowledge of the full solutions to the classical equations of motion is necessary in order to establish whether or not there exists gradual deterioration of the amount of predictability. It is well known, that generically perturbative solutions to the classical equations of motion are valid only for short interval of time. The same argument is more pointedly true in the case of quantum open systems, when one wants to study environmentally induced decoherence and classicality  \cite{Har}. 

\subsection{ Open systems}

So far our discussion has been concentrated on closed systems. When our quantum system is coupled to an environment the evolution is not unitary and a class of interesting phenomena related to predictability appear. Most prominent amongst them is the emergence of superselection rules, namely that some class of environments produce a rapid diagonalisation of the density matrix in some phase space basis.
\par
So, examination of classicality in presence of an environment requires besides the study of predictability preservation a quantification of how close the density matrix is diagonal to a phase space basis ( and what is such a terminology). In this context, the SW entropy has been used before: a lower bound has been computed for a particular class of open systems \cite{HaAn} (see also \cite{An,HaDo} for a consistent histories analysis of the classical behaviour in such systems and \cite{HaAn,HuZh} for other measures of predictability.). 
\par 
Usually the discussion is carried out within the formalism of the Caldeira - Leggett model, where an one - dimensional particle is evolving under a potential $V(Q)$ and in contact with a thermal bath of harmonic oscillators. In the high temperature regime, the corresponding master equation is Markovian and reads
 \begin{eqnarray}
 \frac{\partial \rho}{\partial t} &=& \frac{1}{i \hbar} 
[ \frac{p^2}{2 M} + V(x) , \rho ] 
\nonumber \\ 
&-&\frac{\gamma}{ i \hbar} [x,\{ \rho,p\}] - 
\frac{D}{\hbar^2} [x,[x,\rho]]  
\end{eqnarray}
where $D = 2M \gamma k T$ , $k$ the Boltzmann constant, $\gamma$ a dissipation constant depending on the details of the coupling.
\par
The analysis of the behaviour of this model has been quite thorough, so instead of giving a full treatment we shall restrict ourselves to some remarks that are particularly relevant to our approach and have not been made in the aforementioned references.
\par
One question of relevance is whether the characterisation of predictability for various potentials given in section 3.3 changes by the introduction of a thermal environment. Now for quadratic potentials any coherent state evolves into a Gaussian, the center of which is given by the classical equations of motion (actually there is a stronger statement involving the Gaussian approximation for general potentials \cite{An}, but we shall not need this here.) Hence the relevant object is the density matrix 
\begin{equation}
\rho(x,y) =  \left( \frac{\pi \hbar}{\alpha (1+s)} \right)^{1/2} \exp \left( - \frac{\alpha}{2 \hbar} (x^2 + y^2) - \frac{\alpha s}{\hbar} xy + i \frac{\alpha r}{2 \hbar} (x^2 - y^2) \right)
\end{equation}
where $ 0 \leq s < 1$. Up to a Weyl transformation, this is the most general Gaussian density matrix. It is straightforward to compute the corresponding SW entropy
\begin{equation}
I = 1 + \frac{1}{2} \log \left( \frac{ 4 \alpha (1+s)}{ \frac{1}{4 \sigma^2} + \sigma^2 \alpha^2(1 + s^2 + r^2) + \alpha}   \right)
\end{equation}
Note that the parameter $s$ must lie between $0$ and $1$ in order that the function (3.49) corresponds to a true (positive) density matrix.
 Now, let us consider the case of a harmonic oscillator. It is known that for $t >> \gamma^{-1}$ any initial state approaches exponentially the thermal state , for which  (taking again the natural choice  $ \sigma^2 = (4m \omega)^{-1}$)
\begin{equation}
I \simeq  1 + log \frac{kT}{ 2 \pi \hbar \omega }
\end{equation}
which is the classical Gibbs entropy. This implies that as long as our phase space sampling volume $V$ is much larger than $kT/\omega$ (the size of the thermal fluctuations) one can meaningfully talk about the particle moving according to classical dissipative equations of motion, fluctuations around predictability becoming eventually fully thermal (see \cite{An} for more details). 
\par 
In the free particle case, the interest lies in whether the modification due to the environment is sufficient to cause a reduction in the asymptotical rate of increase of the SW entropy. Using the extensive calculations in \cite{HuZh} we find that asymptotically
\begin{equation}
I \simeq 1 + \log \frac{kTt}{\hbar \gamma} 
\end{equation}
Hence the free particle exhibits again a logarithmic in time increase in its entropy hence essentially destroying the degree of predictability in the same manner as in the no environment case. But the important point, is that the corresponding fluctuations are to be interpreted as thermal (hence classical) rather than fully quantum as in the former case.
\par
Now in the case of an open system, the SW entropy as a measure of fluctuations cannot separate between the ones induced by the environment and the intrinsic to the system itself. What we would like is a quantification of the degree the distribution function of a quantum  open system behaves as a classical one. The key for an answer lies in equation (3.4): the SW entropy is always larger than the von Neumann entropy. The latter enompasses the degree of mixing of the quantum state, hence their difference ought to be a measure of the purely quantum mechanical unperdictability. For the Gaussian density matrix (3.49) the von Neumann entropy reads 
\begin{equation}
S = - \log(1-s) - \frac{s}{1-s} \log s
\end{equation}
 Indeed one can check  that both for the free particle and the harmonic oscillator in the Caldeira- Leggett environment at long times 
\begin{equation}
I - S \simeq 0
\end{equation}
hence the fluctuations around predictability of the particle are asymptotically classical ones. 
\subsubsection{A criterion for pointer basis}
It is often stated in the bibliography that coherent sttes are essentially the pointer basis to which a density matrix becomes diagonal, due to interaction with the environment, these being the natural choice of phase space localised states.
But some caution should be exercised on that point. A large class of density matric can be diagonalised in a coherent basis, the latter being overcomplete. The requirement is essentially the existence of the P-symbol $f({\bf q},{\bf p})$ given by
\begin{equation}
\hat{\rho} = \int \frac{d {\bf p} d{\bf q}}{(2 \pi \hbar)^n} f({\bf q},{\bf p}) |{\bf pq} \rangle \langle {\bf pq}| 
\end{equation}
Now, recall the property (2.7) of quantum mechanical information. If a density matrix is diagonal in a given basis, the corresponding information is equal to the von Neumann entropy. This can provide a criterion for determining the pointer basis. Indeed, consider the SW entropy defined with respect to a particular coherent state family, labeled by the defining vector $\xi$. One 's task should be then to determine the family by requiring the minimisation of $I_{\xi} - S$. 
\par
One can use a result due to Wehrl to improve the characterisation. If the P-symbol of a density matrix exists and is positive then there exists a lower bound to the von Neumann entropy
\begin{equation}
S \geq - \int \frac{d {\bf p} d{\bf q}}{(2 \pi \hbar)^n} f({\bf q},{\bf p}) \log f({\bf q},{\bf p}) = I_P
\end{equation}
Hence the quantity $I - I_P$ whenever $I_P$ exists is an upper bound to $I - S$ characterising the pointer basis. Now, if a P symbol is positive then its distance in norm (determined by the coherent state metric from the Q symbol is of the order of $\hbar$). Hence, a sufficient criterion for the determination of the pointer basis is the P- symbol corresponding to that basis becoming positive rapidly for all choices of initial states. Such a basis has been constructed in \cite{HaZo}, using ideas from the quantum state diffusion picture of quantum open systems, and it consists of Gaussian states with small value of the squeezing parameter.
\par
We should remark at this point, that information theoretic criteria seem to  be strong enough to discuss the issue of pointer basis, without referring to the notion of the reduced density matrix. For instance in a combined system living in a Hilbert space $H_1 \otimes H_2$ , we could verify that the system $1$ gets asymptotically diagonalised in the basis of the operator $\hat{A}$ when the quantity $I_{A \otimes 1} - S$ is close to zero. This could also generalise to the case where there is no natural splitting between system and environment, hence the reduced density matrix is not naturally defined. One then consider the information associated to some self-adjoint operator $\hat{A}$, typically with degenerate spectrum so that a degree of coarse graining is to be incorporated, and compare it to the von Neumann entropy. If technically feasible, this would provide an alternative way of checking, for instance, the classicalisation of hydrodynamic variables or of variables corresponding to Boltzmann-type of coarse graining \cite{CaHu2}.
\par
Indeed this construction might easily be seen in the context of phase space classicality. One can construct a lattice on phase space, consisting of,say, cubic cells $C_i$ with volume much larger than $\hbar$ and then consider the operator $\hat{A} = \sum_i \lambda_i \hat{P}_{C_i}$, where $\lambda_i$ are real numbers and $P_{C_i}$ the relevant quasiprojectors. Using the properties of quasiprojectors it is an easy task to verify that the corresponding information $I_A$ isgenerically  of the order of $\epsilon$ for phase space localised states , $\epsilon$ being the classicality function of the quasiprojectors.
\par
For  similar ideas, using the von Neumann entropy to identify the most stable  states in evolution under an environment, the reader is referred to \cite{Zur}.
\section{Field theory}
In this section, we shall try to examine whether our results can be generalised to a field theory case. A quantum field, being a system with infinite number of degrees of freedom (an infinite dimensional phase space) is expected to have much more complicated behaviour. In the case of interacting fields, the study of classical behaviour is much more complicated, since as we discussed earlier the notion of predictability is a non-perturbative phenomenon.
\par
The SW entropy is expected to play again an important role for the identification of classical predictability. But we should note, that a quantum field is itself a thermodynamical system (due to its infinite number of degrees of freedom), hence it would be important to see whether the SW entropy is connected to its proper thermodynamical entropy. It would be indeed an appealing picture, if we could (even in the simple free field case) transfer the notion of entropy due to mixing in phase space also in the field theory case.

\subsection{The notion of classicality in field theory}

A possible divergence between quantum field theory (QFT) and quantum particle mechanics, as far as the issue of classicality is concerned, lies mainly on the facts that \\
1. QFT describes a system with infinite number of degrees of freedom \\
2. QFT is relativistically invariant. \\
\\
The queston then arises, whether these differences are sufficient to necessitate a different approach towards the issue of classicality. Again, we are going to concentrate on the notion of Hamiltonian classicality, i.e. whether and in what regime QFT behaves as a classical field theory. The fact that we have a system with infinite number of degrees of freedom, necessitates the consideration of other type of quasiclassical domains associated to the field's thermodynamic or hydrodynamic behaviour. We shall return to this issue later, but for now we shall concentrate on the possible emergence of a classical field theory.
\par
The condition for classicality we developed in section 3, is at first sight sufficiently general to encompass the case of QFT as well. It makes no reference to whether the phase space is finite or infinite dimensional. But the issue of integration in an infinite dimensional space is quite complicated and there is no apparent way of how one would construct a classicality parameter associated to each quasiprojector, that would have an intuitive geometric meaning. Indeed in an attempt to generalise Omn\'es theorem for the case of free fields Blencowe \cite{Blen} restricted the consideration to finite dimensional phase space cells. Such a restriction implies a consideration of essentially a finite number of modes. While this might give sufficient physical information for free fields (studying the modes is standard practice for instance in cosmological setting) clearly cannot be transferred to the non - linear case.
\par
On the other hand, the function of the SW entropy as a measure of predictability seems to be unaffected by the transition to the infinite dimensional case. Indeed, one can define coherent states for the fields (we shall give the basic conventions later) even in the interacting case and there is a well defined notion of  integration over the infinite dimensional phase space.
\par
The Hilbert space of a quantum field carries a unitary represenatation of tha canonical commutation relations 
\begin{equation}
[\hat{\Phi}(x),\hat{\Pi}(x')] = i \delta (x,x')
\end{equation}
where $x$ and $x'$ are points on a Cauchy surface $\Sigma$.
We  can define the field coherent states as
\begin{equation}
|w \rangle = | \phi, \pi \rangle = e^{ i \Phi(\pi) + i \Pi(\phi)} |0 \rangle
\end{equation}
The relation between the complex function $w(x)$ (an element of $L^2(\Sigma)$ and the phase space coordinates $\phi$ and $\pi$ is dependent on the choice of the representation. Now, if the Hamiltonian is quadratic the vacuum state is a Gaussian (in either the Schr\"oddinger or the Bargmann representation) and so are our coherent states.
\par
Given then a density matrix $\rho$ one can define the probability distribution 
\begin{equation}
p(w) = \langle w| \rho |w \rangle / \langle w|w \rangle
\end{equation}
and from this define the SW entropy as in equation (3.1), where now the integral measure is $Dw Dw^*$, which is the well defined Gaussian integral on the field phase space.
\par
Note, that for the free fields the Gaussian nature of coherent states reproduces again the lower bound (3.5) for SW entropy, but in the case of interacting fields this is not any more true (the vacuum is not a Gaussian). Also in the interacting field's case, it  makes no sense to consider Gaussian coherent states, for they generically do not exist in the field 's Hilbert space. This marks a significance difference from the particle QM case where one could always consider and study the SW entropy minimising Gaussian coherent states, 
\par
For technical reasons therefore we shall be forced to concnentrate only on the free field case. In the case of Minkowski spacetime, time evolution with the free Hamiltonian is rather trivial. The coherent states are preserved, and the analysis proceeds as in the simple harmonic oscillator case. There is no SW entropy production and a classical state will remain classical even as time increases. The same can be shown to be true in the presence of an external source coupled linearly to the field.
\par
But more interesting is the case of a field in a curved dynamical spacetime. These cases are relevant in the cosmological context, and to their examination we shall return shortly.

\subsection{Field theory in cosmological spacetimes}
The evolution of the vacuum states for a field in a time dependent cosmological spacetime essentially corresponds to  a linear canonical transformation acting on the field. Hence equation (3.23) for the SW entropy is applicable here, provided that the trace exists, since of now $A$ and $B$ are operators in an infinite dimensional Hilbert space. Actually this condition is equivalent to $Tr \bar{K}K < \infty$ and since $A$ is bounded equivalent to $Tr B^{\dagger}B < \infty$. THis is of course  the necessary condition for the Bogolubov transformation to be unitarily implementable or the total number of created particles to be finite.
\par
In cosmological situations (or at least in the models usually employed) this is not the case , but still one can define a kind of entropy density by restrict ing the spatial integration involved in the trace to a finite region, dividing by its volume and in the end taking the latter into infinity.
\par
It is usually the case that the Bogolubov transformation couple only a finite number of modes, in which case it is meaningful to define an entropy per particle by concentrating on the relevant finite dimensional subspaces.
\par
A case which has been explicitly discussed is the case where the Bogolubov transformation break into two dimensional blocks involving the modes labeled by ${\bf k}$ and $- {\bf k}$, in each block the transformation given by a two dimensional squeezing transformation. Transformations of  type (3.26) appears for instance in pair creation of gravitons (or scalar fields) from the vacuum. 
\par
An important point in this case is that the squeezing parameter $r_{{\bf k}}$ is related to the   number of created particles on mode ${\bf k}$, $n_{{\bf k}}$ by
\begin{equation}
n_{{\bf k}} = \sinh^2 r_{{\bf k}}
\end{equation}
hence the SW entropy per mode can be written 
\begin{equation}
I_{{\bf k}} = 1 + \log (1 +n_{{\bf k}})
\end{equation}
and the entropy  increases with the number of particles created.
In general the knowledge of the Bogolubov coefficients in any cosmological model enables us to straightfowardly compute the SW entropy. Such calculations have been done in a number  of cases \cite{Gasp,Ro,Bran} and is not the point we intend to  pursue here. We are rather more interested in some interpretational issues.
\paragraph{Field classicalisation:} As we have argued in the previous chapter, the SW entropy (or rather the relative SW entropy) is a measure of the deviation of the system from classical deterministic behaviour, while the quantity $I - S$ is a measure of the deviation from classical stochastic behaviour. Given the fact that in most relevant models the squeezing parameters increase with time ( for conformally coupled massless scalar field in de Sitter spacetime $r_{{\bf k}} = H t$) we conclude that rather than producing classicality, time evolution in time dependent background enhances non-classical behaviour. We have given detailed argumentation in the previous section, but we should also examine a number of possible counterarguments.
\par
We have already examined the case the argument that extreme squeezing in one direction is essentially equivalent to diagonalisation of the state in some pointer basis. We argued against this by pointing that classicality and determinism is essentially a phase space issue. Still, one can argue \cite{KiPo2} that in an operational sense the highly squeezed states corrspond to classical states , in the sense that the observationally relevant quantities are field amplitudes rather than field momenta. Setting aside the measurement-theoretic truth of this assertion, we should point out that this notion of classicality is not robust to even small external perturbations.
 This can be seen  even in non - relativistic quantum mechanics from inspection of equations (3.39) and (3.40). Any interaction terms couple intrinsically position and momentum uncertainties and are prone to increase the small position uncertainty of a squeezed state. I
In addition it is not robust in the presence of a decohering environment. Even when the system couples to the environment via its configuration space variables, the pointer basis is not the position basis but of a coherent state type. This has been demonstrated  in \cite{Zur, HaZo}.
\par
 Another argument usually put forward is that at the limit of large squeezing the number of created particles becomes very large and hence can be taken in some sense to correspond to classical behaviour. The problem with this, is that {\it a priori}  classicality is insensitive to the number of particles (one can easily construct {\it Schr\"odinger cat states} even for a many particle system and there is no guararantee from first principles , unless some explicit mechanism being described, that suppresses such interferences. What is more, as is known from quantum optics 
 the distribution function of photons  in squeezed states is highly non classical (non - Poisson) \cite{OOO} . 
\par
Given the fact that field classicalisation is important in any discussion of inflation, one should start examining alternatives. Coupling of the fields to an environment might seem to provide a solution to the problem, turning the quantum fluctuations into thermal ones, and indeed seems quite probable. But still one has to show that classicality does appear in such system. According to our argumentatation the calculation of $I -S$ is a good guide for obtaining a classical stochastic process. But still there are some problems. First of all the difficulty of separating between system and environment (in a non linear theory this splitting seems to be quite arbitrary \cite{CaHu2, An2}). Second, we should not forget that even the environment undergoes squeezing due to the time dependence of the scale factor and there is no guarantee (at least not from the well studied examples ilike the Caldeira Leggett model) whether such feature might render clasicalisation problematic. An investigation of this issue will be taken elsewhere. 
\par
Another possibility, of the classicalisation of much more coarse - grained hydrodynamic (rather than phase space quantities as discussed here) is tentatively discussed in \cite{An3}.
\paragraph{Phenomenological entropy}
The other important question is whether the SW entropy for the fields can be taken to represent the phenomenological entropy of the matter as defined in the latter universe. This has been argued in reference \cite{Gasp}. This would be indeed an appealing feature, since the SW entropy can be conveniently interpreted as a measure of the phase space mixing induced to the field by the classical evolution.
\par
What entropy actually corresponds to the phenomenological thermodynamic entropy is often a difficult quastion to answer. In standard equilibrium thermodynamics the von Neumann entropy of a thermal density matrix is to be identified with the thermodynamic entropy, by the consideration of a quantum mechanical version of the macrocanonical distribution, implicitly aknowledging the openness of the thermodynamic system as coupled to a heat bath. 
\par
In the case of cosmology our system is essentially closed and far from equilibrium. It seems therefore that the entropy ought to be identified with some coarse grained version of von Neumann entropy. The SW entropy is such a candidate, involving minimal smearing over phase space and being very close to Gibbs entropy, but is not the only one. Any thermodynamic description necessitates the identification of a finite number of macroscopic degrees of freedom describing the system. Should we wish for such a description in a quantum field, we ought to perform definitely further coarse graining, as for instance focusing on a set of hydrodynamic variables characterising it , or tracing out the effect of higher order correlation functions, smearing over spatial or spacetime regions etc.
\par
The point we want to make is that a thermodynamical description has to be given in terms of essentially classically behaving quantities. This is not the case of the minimally coarse grained phase space description implied by the SW entropy. It seems therefore necessary that extra coarse graining would be necessary in order to obtain a quantity that could naturally be considered as the thermodynamical entropy. For these reasons we are rather reluctant to consider the SW entropy as a measure of the actual thermodynamical entropy of the quantum field, and we are restricted to its interpretation as a measure of deviation of classicality and phase space mixing due to time evolution.

\section{Conclusions}

To conclude, we would like to put our results in a different perspective, that might turn out to provide an alternative way to discuss the issue of classicality.
\par
One can use coherent states to define unequal time n-point functions on phase space for any quantum systems (see for instance \cite{Sri} and references therein). Such objects, provided they satisfy the Kolmogorov conditions, can be used to define a measure on phase-space paths and hence a stochastic process. As expected from the Bell - Wigner theorem this is not true in the case of quantum mechanics. But then the question arises, when is the quantum process close to a classical process and how do we quantify the notion of closeness?
\par
The quantity $I - S$ we examined in this paper is able to play this role. This having value of the order of unity is a sign that the quantum mechanical evolution can be approximated by a classical stochastic process. Of course, classical determinism cannot be seen from the inspection of this quantity: the evolution of a superposition of two phase space localised states in presence of decohering environment is such an example: the system behaves classically, but stochastically rather than deterministically.
\par
We can easily see that our criterion for a classical state, corresponds to this way of addressing classicality. Indeed, given an initial density matrix and the evolution equation, the ``quantum stochastic process'' describing the system in phase space is uniquely constructed.  As argued, the quantity $I - S$ can provide a good  quantifying criterion for phase space classicality, giving a single quantification even for system with large number of degrees of freedom. But of course, a more complete and satisfactory description, would be given by translating our stated classicality criterion into a stochastic process language. This issue is currently our main investigation.

\section{Aknowledgements}
I would like to thank J.J. Halliwell, A. Zoupas and A. Roura for discussions and comments. The research was supported by Generalitat de Catalunya by grant 96SGR-48.

\end{document}